\documentclass[aip,jcp,reprint]{revtex4-1}
\usepackage{graphicx}% Include figure files

\usepackage{amsfonts, amsmath, amssymb,latexsym}

\usepackage{dcolumn}% Align table columns on decimal point

\usepackage{bm}% bold math

\newcommand{\me}{\mathrm{e}}

\newcommand{\dif}{\mathrm{d}}

\begin{document}

\title{Loop-Closure and Gaussian Models of Collective 
Structural Characteristics of Capped PEO Oligomers in Water}

%\author{}
%\email{mchaudha@tulane.edu}
\author{M. I. Chaudhari, L. R. Pratt} 
%\email{lpratt@tulane.edu}
\affiliation{Department of Chemical and Biomolecular Engineering, Tulane University, New Orleans, LA 70118}
\author{M. E. Paulaitis} 
\affiliation{Department of Chemical and Biomolecular Engineering, The Ohio State University,
Columbus, OH 43210}

%\email{}
%\affiliation{}

\date{\today}

\begin{abstract} Parallel-tempering MD results for a
CH$_3$(CH$_2$-O-CH$_2$)$_m$CH$_3$ chain in water are exploited as a data-base
for analysis of collective structural characteristics of the PEO globule with a
goal of defining models permitting statistical thermodynamic analysis of
dispersants of Corexit type. The chain structure factor, relevant to neutron
scattering from a deuterated chain in neutral water, is considered specifically.
The traditional continuum-Gaussian structure factor is inconsistent with the
simple $k \rightarrow \infty$ behavior, but we consider a discrete-Gaussian
model that does achieve that consistency. Shifting-and-scaling the
discrete-Gaussian model helps to identify the low-$k$ to high-$k$ transition
near $k \approx 2\pi/0.6~\mathrm{nm}$ when an empirically matched number of
Gaussian links is about one-third of the total number of effective-atom sites.
This short distance-scale boundary of 0.6~nm is directly verified with the
$r$-space distributions, and this distance is thus identified with a natural
size for coarsened monomers. The probability distribution of $R_g{}^2$ is
compared with the classic predictions for both Gaussian model and freely-jointed
chains. $\left\langle R_g{}^2(j)\right\rangle$, the contribution of the $j$-th
chain segment to $\left\langle R_g{}^2\right\rangle$, depends on contour index
about as expected for Gaussian chains despite significant quantitative
discrepancies which express the swelling of these chains in water. Monomers
central to the chain contour occupy the center of the chain globule. The density
profiles of chain segments relative to their center of mass can show distinctive
density structuring for smaller chains due close proximity of central elements
to the globule center. But that density structuring washes-out for longer chains
where many chain elements additively contribute to the density profiles.
Gaussian chain models thus become more satisfactory for the density profiles for
longer chains. 

\end{abstract}

\maketitle

\section{Introduction} Arguably the most important water-soluble synthetic
polymers,\cite{Alessi:2005ix,Norman:2007kq} (-CH$_2$-O-CH$_2$-)$_n$ chain
molecules are intrinsic to the dispersant materials applied to oil spills,
\cite{dispersants} and can be soluble also in organic solvents. With -H (and
thus methyl -CH$_3$) ends polyethylene oxide (PEO) is a common name, and we will
use that name generically when the chains have arbitrary capping groups. With
hydroxyl -OH terminations these chains are typically called polyethylene glycol
(PEG), and we will use that name in discussing experiments that study that case
specifically.

For dispersants used on oil spills,\cite{dispersants} and for other
applications,\cite{Lin:2012dz} PEO chains are often decorated with junctions or
tails or caps. Correlations associated with capping groups then focus molecular
structural analyses, molecular specificity in understanding loop closure being
an outstanding interest.\cite{Weikl:2008ii} Those correlations can be the
targets of neutron scattering experiments\cite{chaudhari_communication:_2010}
with deuterated chains, under the important limitation of chemical feasibility
of specific isotopic labeling for the caps. We have studied
X(-CH$_2$-O-CH$_2$-)$_n$X with a variety of capping groups X for just those
reasons.\cite{MICThesis} Capping groups can non-trivially change solution
properties, particularly for the short-chain oligomers, and those changes have
been of specific interest.\cite{Dormidontova:2004uz}  

Here we analyze the X = CH$_3$ case. With this capping choice, neutron
scattering experiments also characterize C-C contacts for C atoms closer along
the chain contour than the end-caps, \emph{i.e.,} short-circuited loops
contribute to those neutron scattering results. We utilize simulation results
theoretically to investigate anticipated neutron scattering studies that
interrogate loop-closure. We consider a structure factor model that spans
small-angle and diffraction regimes, which thus highlights the molecular-scale
features that might be sought.

The structure factor model that we analyze assumes ideal Gaussian chains, and is
particularly simple. Nevertheless, it is more complicated than the most
traditional continuum-Gaussian model, and the distinction is essential for the
success of the model in these comparisons.

We then further test other aspects of Gaussian chain models against simulation
results. Since Gaussian chain models are particularly simple, these alternative
aspects are typically collective characteristics. Though limited in molecular
detail, collective characteristics have countervailing advantages of wide
utility. Initial examples include the probability distribution, $P(R_g{}^2)$, of
the square radius of gyration, and then the $\left\langle
R_g{}^2(j)\right\rangle$ for the $j$-th chain segment, contributing to the
decomposition \begin{eqnarray} \left\langle R_g{}^2\right\rangle =
\frac{1}{n+1}\sum_{j=0}^{n} \left\langle R_g{}^2(j)\right\rangle~.
\label{eq:rg2decomposition} \end{eqnarray} 

Building from $\left\langle R_g{}^2(j)\right\rangle$ results, we consider
density profiles of chain globules described by a Gaussian chain model. Our goal
is to establish simple models that permit statistical thermodynamic evaluation
of the surface tensions of aqueous electrolyte solutions with hydrocarbon
liquids\cite{nichols1982disentanglement,Wilson:1984cs,nichols1984salt,pratt1992contact}
when dispersant materials are deployed. As an example of a specific
characteristic that should be helpful, we note that the dielectric constant of
aqueous PEO solutions depends linearly on the water volume
fraction.\cite{Borodin:2002gx,MICThesis}

These systems are notorious for physical complexity despite their chemical
simplicity.\cite{Israelachvili:1997tr} But a broad physical description of these solutions is that
water is a good solvent for PEO chains, which are swollen by the solvent. Our
results for the osmotic second virial coefficients for
CH$_3$(CH$_2$-O-CH$_2$)$_{11}$CH$_3$, $B_2 >0$ obtained elsewhere from
multi-chain solution simulations,\cite{MICThesis} indicates repulsive
inter-segment interactions at ambient $\left(T,p\right)$ conditions. For an
experimental perspective on PEG osmotic pressures, see Cohen, \emph{et
al.}\cite{Cohen:2009ct,Cohen:2012ul} Consistent with repulsive inter-segment
interactions, we find $\left\langle R_g{}^2\right\rangle \propto m^{1.3}$. Of
course, that exponent was not the goal of the calculations implemented, which
are detailed below.

Nevertheless, the solution environment can sensitively affects PEG
conformations.\cite{Alessi:2005ix,Norman:2007kq} PEG molecules are helical in
$n$-propanoic, isobutyric, and isopentanoic acid solutions coexisting with
liquid water,\cite{Norman:2007kq} with helix formation requiring a trace of
water.\cite{Norman:2007kq} In contrast, these chains are generically coiled in
aqueous solutions and also in acetic acid, isobutanol, and $n$-butanol.
Conformational sensitivity is associated with size fractionation of PEGs between
water and isobutyric acid.\cite{Alessi:2005ix,Norman:2007kq} 

The versatility of PEG polymers solutions makes them a challenge for molecular
thermodynamics. Flory-Huggins interaction parameters, experimentally evaluated,
show substantial but different \emph{composition} dependences for PEG in water
and methanol,\cite{Bae:1993uj,ZafaraniMoattar:2006kr} but in ethanol only minor
dependence on composition.\cite{ZafaraniMoattar:2008tb}

The results below extend the aqueous solution calculations discussed in a
preliminary report that compared $n$-hexane solvent with
water.\cite{Chaudhari2014} Previous simulation
calculations\cite{Borodin:2001cn,Borodin:2002gx,Lee:2008hb,Choi:2013dg,Mondal:2014gp,Starovoytov:2011km}
evaluate different aspects of these solutions, and  give a helpful baseline on
which the present modeling builds.

\section{Methods} The simulation calculations below treated a single
CH$_3$(CH$_2$-O-CH$_2$)$_m$CH$_3$ molecule in water by molecular dynamics, using
parallel tempering\cite{Earl:2005fv} to achieve enhanced sampling of chain
conformations. We evaluated system sizes of $N_{\mathrm{water}}$ = 1000 ($m$ =
11), and $N_{\mathrm{water}}$ = 2000 ($m$ = 21, 31). The chain molecules were
represented by optimized potentials for liquid simulations
(OPLS-AA),\cite{Jorgensen:1996vx} and the SPC/E model was adopted for
water\cite{Berendsen:1987uu} implemented with the GROMACS 4.5.3 molecular
dynamics simulation package.\cite{vanderSpoel:2005hz} Long-range electrostatic
interactions were treated in standard periodic boundary conditions using
particle mesh Ewald with a cutoff of 0.9~nm. The Nos\'{e}-Hoover thermostat
maintained the temperature and hydrogen atom bond-lengths were constrained by
the LINCS algorithm. After energy minimization, density equilibration with
$\left(T=300\mathrm{K},p=1~\mathrm{atm}\right)$ MD calculations established the
constant volumes for each parallel tempering simulation. The parallel tempering
spanned the 256-550K temperature range with 32 replicas (for $m$ = 11, and 21
cases) and 40 replicas (for $m$ = 31). Parallel tempering swaps were attempted
at a rate of 100/ns, and the temperature grid resulted in a success rates of
15-25\%. Production calculations for each replica set were extended to 10 ns.

\begin{figure}[h]
\begin{center} \includegraphics[width=3.2in]{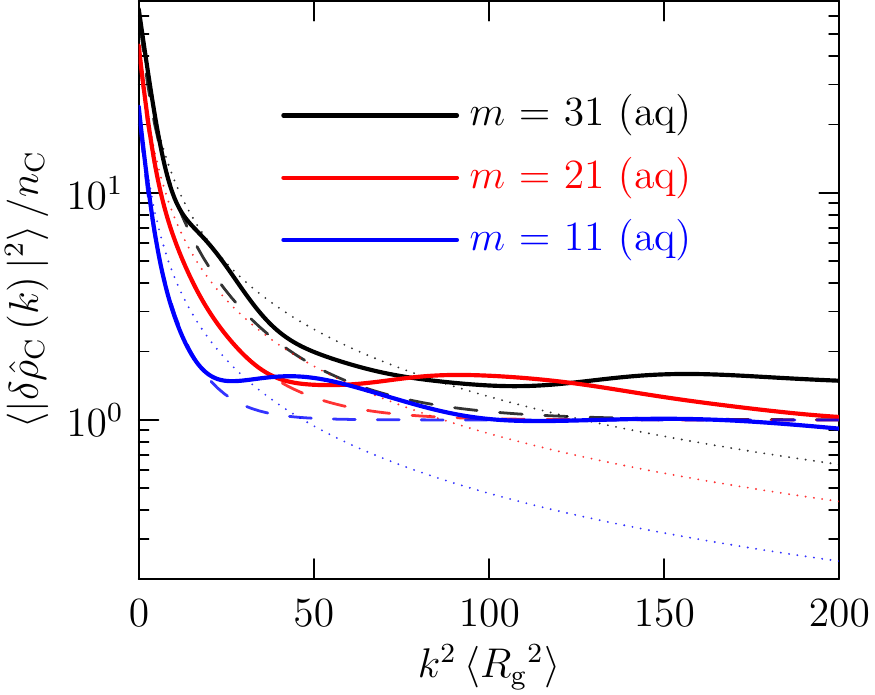}
\end{center}
\caption{The traditional continuum-Gaussian structure factor
(Eq.~\eqref{eq:Debye}, dotted curves) contrasted with a discrete-Gaussian model,
shifted and scaled, which is consistent with the correct $k \rightarrow \infty$
limit (Eq.~\eqref{eq:shiftandscale}, dashed curves). The direct numerical
simulation results are the solid curves. For the model
Eq.~\eqref{eq:shiftandscale}, the number of Gaussian links, $n$, was adjusted
for agreement throughout a low-$k$ regime. In all cases here $n$ was about a
third of the number of heavy atoms of CH$_3$(CH$_2$-O-CH$_2$)$_m$CH$_3$,
\emph{i.e.,} in coarse-grained models of these oligomers, coarsened monomers can
represent about three effective-atoms for this characteristic. The local maxima
for $k>0$ here correspond to $k \gtrsim 2\pi/0.6\mathrm{nm}$. \label{fig:fig5}} \end{figure}

\section{Results and Discussion}
The structure factor 
\begin{equation}
\left\langle \vert \delta \hat{\rho}_\mathrm{C}\left( k \right)\vert^2 
\right\rangle/n_\mathrm{C} = 1 + \int
\left(\frac{\sin kr}{kr}\right) 
\left\langle \rho_\mathrm{C}\left(r\right)\vert 0
\right\rangle
\dif \mathbf{r}  ~,
\label{eq:sk}
\end{equation}
addresses CC loop-closure contacts comprehensively, in contrast  to
chain-end closure exclusively which would be  targeted by labelled
ends.\cite{chaudhari_communication:_2010} Here $n_\mathrm{C}$ is the number of
C-atoms in the solution, and $\left\langle \rho_\mathrm{C}\left(r\right)\vert 0
\right\rangle$ is the density, conditional on placement of a C-atom at the
origin, of other C-atoms at radius $r$. Since our calculations here treat only one
chain molecule, $\left\langle \rho_\mathrm{C}\left(r\right)\vert 0
\right\rangle$ is the density of other, \emph{intra}molecular  C-atoms, and is
normalized to one less than the number of C-atoms in a solute chain; in our case
\begin{equation}
\left\langle \vert \delta \hat{\rho}_\mathrm{C}\left( 0 \right)\vert^2 
\right\rangle/n_\mathrm{C} =2\left(m + 1\right) ~.
\label{eq:zeroklimit}
\end{equation}
Inverse to Eq.~\eqref{eq:sk} is
\begin{multline}
\left(\frac{1}{2\pi}\right)^3
		\int 
	\left(\frac{\sin kr}{kr}\right)
	\left(\left\langle
\vert \delta \hat\rho_\mathrm{C}\left(k\right)\vert^2
\right\rangle/n_\mathrm{C}\right) \dif \mathbf{k} = \\
\delta\left(\mathbf{r}\right) + \left\langle 
\rho_\mathrm{C}\left(r\right)\vert 0
\right\rangle~.
\label{eq:inverse}
\end{multline}
The simple result
\begin{multline}
\left\langle
\vert \delta \hat\rho_\mathrm{C}\left(k\right)\vert^2
\right\rangle/n_\mathrm{C} = \left(2 m+1\right) \\
\times 
\left\lbrack
\exp\left(-k^2\left\langle R_g{}^2\right\rangle\right) - 1 + k^2\left\langle 
R_g{}^2\right\rangle\right\rbrack  \\
\times \frac{2}{\left(k^2\left\langle R_g{}^2\right\rangle\right)^2}~.
\label{eq:Debye}
\end{multline}
is the continuum-Gaussian model that we consider.\cite{BernePecora} This satisfies the anticipated 
low-$k$ behavior, but
not the $k \rightarrow \infty$ value associated
with the $\delta\left(\mathbf{r}\right)$ function of
Eq.~\eqref{eq:inverse}. The result for a discrete-Gaussian chain is
\begin{multline}
\left[
\left\langle
\vert \delta \hat\rho_\mathrm{C}\left(k\right)\vert^2
\right\rangle/n_\mathrm{C}\right]_{\mathrm{DG}} = \\
\left[
\me^{2\zeta/n}\left(n+1\right) - 2\left(\me^{\zeta/n}-\me^{-\zeta}\right) - 
\left(n+1\right)
\right]
\\
/\left\lbrack\left(n+1\right)\left(\me^{\zeta/n} - 1\right)^2\right\rbrack
~,
\label{eq:model}
\end{multline}
where $\zeta$ = $k^2\left\langle R_g{}^2\right\rangle$, and $n$ is the number of 
\emph{Gaussian links}. This has the expected
$\zeta \rightarrow 0$ limit, namely, $\left(n+1\right)$ (Eq.~\eqref{eq:zeroklimit}).  
Then $n \rightarrow \infty$, with
$\zeta$ fixed, leads to Eq.~\eqref{eq:Debye} and clarifies
``continuum'' here.

\begin{figure}[h]
\begin{center} \includegraphics[width=3.2in]{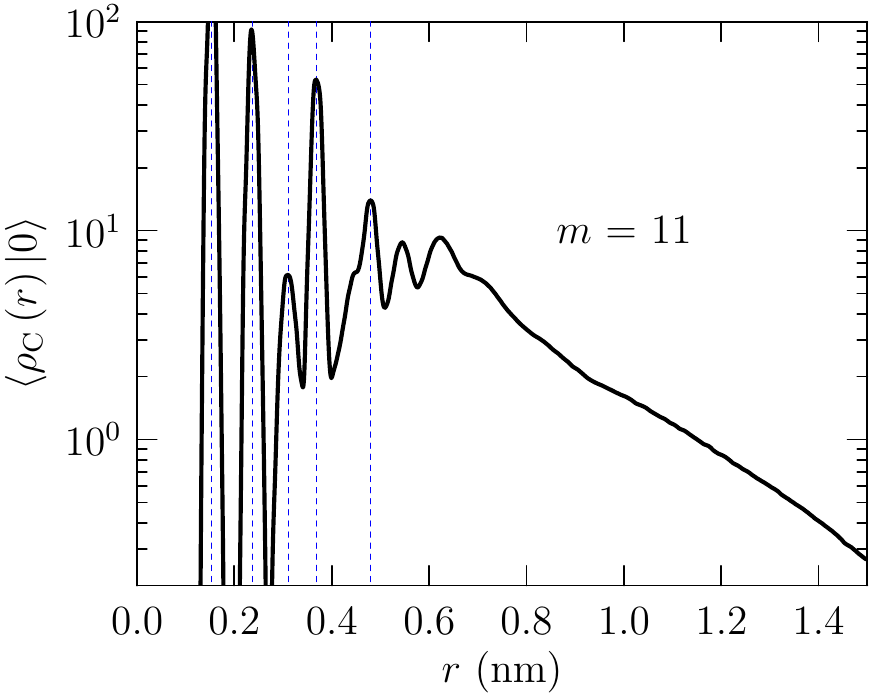}
\end{center}
\caption{Density, conditional on placement of a C-atom at the origin, of other
C-atoms at radius $r$, see text. Consistent with FIG.~\ref{fig:fig5}, the
short-distance regime begins with $r < 0.6$~nm (high-angles for $k >
2\pi/0.6$~nm). The dashed vertical lines indicated distances of specific
interest. The $r\approx 0.38$~nm  peak represents direct, non-bonded CC
contacts that are of interest to investigations of hydrophobic
interactions.\cite{chaudhari_communication:_2010}\label{fig:gSCCm11}}
\end{figure}

We shift and scale the discrete-Gaussian model,  
\begin{multline}
\left\langle
\vert \delta \hat\rho_\mathrm{C}\left(k\right)\vert^2
\right\rangle/n_\mathrm{C}- 1 = \\
\left(\frac{2 m+1  }{n}\right)\left\lbrace\left[
\left\langle
\vert \delta \hat\rho_\mathrm{C}\left(k\right)\vert^2
\right\rangle/n_\mathrm{C}\right]_{\mathrm{DG}}-1\right\rbrace~,
\label{eq:shiftandscale}
\end{multline}
to compare with the simulation data (FIG.~\ref{fig:fig5}). The discrete-Gaussian
model matches the data through the low-$k$ regime more effectively than
does the continuum model Eq.~\eqref{eq:Debye}. Beyond that low-$k$ regime, the
data deviate from the discrete-Gaussian model positively through a local maximum
indicating a short length scale $2\pi/k_\mathrm{max} \lesssim$ 0.6~nm.

We can directly turn to $\left\langle \rho_\mathrm{C}\left(r\right)\vert
0\right\rangle$ for confirmation of this inference (FIG.~\ref{fig:gSCCm11}).
Indeed, the short-distance regime begins with $r < 0.6$~nm, consistent with
identification of high-angles for $k > 2\pi/0.6$~nm (FIG.~\ref{fig:fig5}).

\begin{figure}[h] \begin{center}
\includegraphics[width=3.2in]{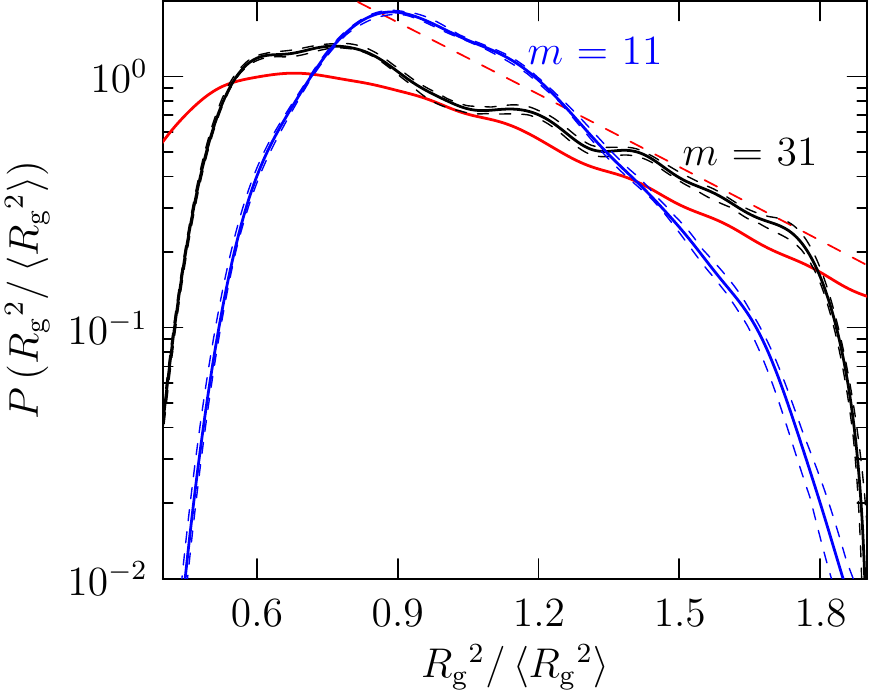} \end{center} 
\caption{The solid red curve is the result for the $m$ = 31 case of a freely
jointed chain, obtained numerically by a straightforward Monte Carlo
calculation; the dashed red curve is the asymptotic
$R_\mathrm{g}{}^2/\left\langle R_\mathrm{g}{}^2\right\rangle \sim \infty$ result
for an ideal Gaussian model,\cite{YamakawaB} close to a simple Gaussian
function. The dashed lines bracket the 95\% confidence intervals approximated by
a bootstrap method.\label{fig:fig1}} 
\end{figure}

In addition to the assumption of ideal behavior for the $n$ coarsened monomers,
the model tested above obviously utilizes an empirical $\left\langle
R_g{}^2\right\rangle$. We next consider the distribution of $R_g{}^2$ for these
chains, compared to Gaussian model results (FIG.~\ref{fig:fig1}). The
distinctions deriving from molecular-scale resolution, cutoffs at minimum and
maximum lengths, are prominent.

\begin{figure}[h] \begin{center} \includegraphics[width=3.2in]{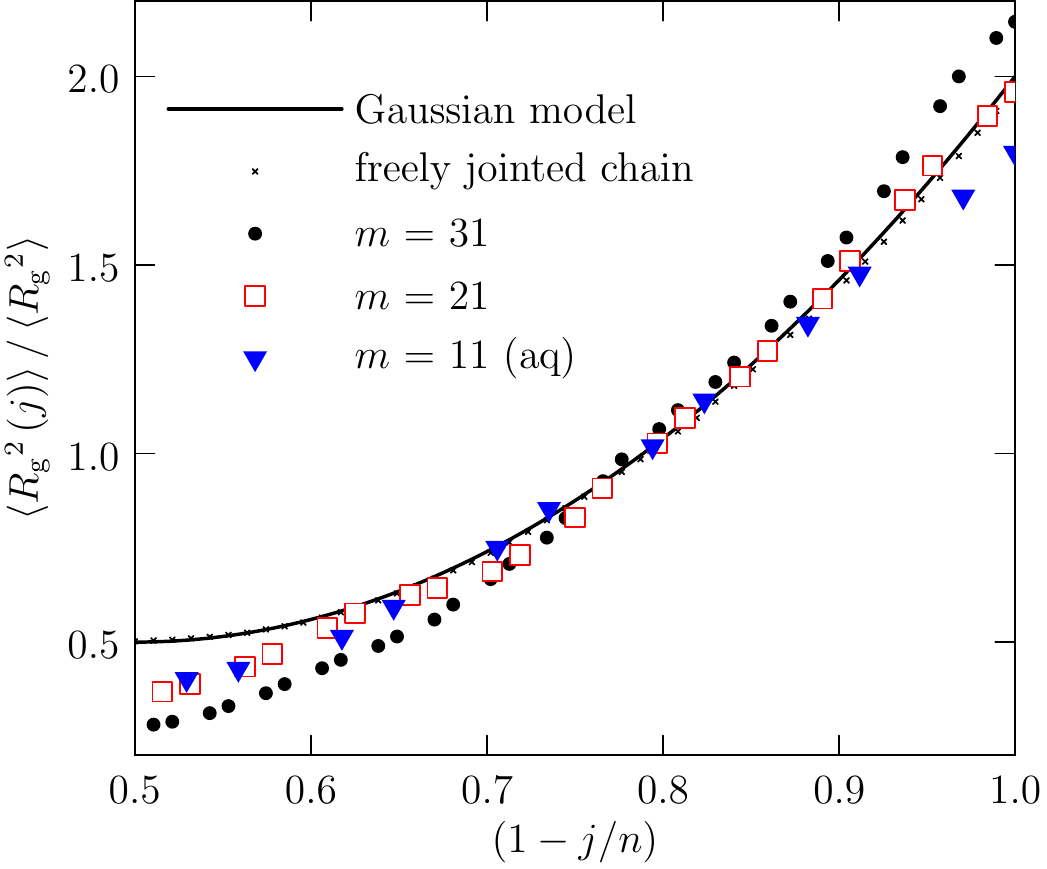}
\end{center} 
\caption{The number of bonds along the heavy atom chain contour is indexed by
$n$, \emph{i.e.} $n$ = 34 which has 35 heavy atoms for the $m$ = 11 chain. The
chain molecule length index $m$ is defined on the basis of the molecular formula
CH$_3$(CH$_2$-O-CH$_2$)$_m$CH$_3$. The solid curve is the function
$2\left\lbrack1-3\left(j/n\right)\left(1-j/n\right)\right\rbrack$ appropriate
for an ideal Gaussian chain.\cite{YamakawaA} The crosses are the results for the
$m$ = 31 case of a freely jointed chain, obtained numerically by a
straightforward Monte Carlo calculation. The right-most triangle corresponds to
a methyl C atom, here index $j$=34; $j$=0 is chemically equivalent. The third
triangle from the right boundary corresponds to $j$=31, chemically equivalent to
$j$ = 3.\label{fig:fig2}} \end{figure}

The comparison (FIG.~\ref{fig:fig2}) of the observed dependence of $\left\langle
R_g{}^2(j)\right\rangle$ on contour index $j$ with that for ideal Gaussian
models shows encouraging agreement. On the otherhand, the discrepancies of ideal
Gaussian behavior from the observed results are much larger than the difference
of the ideal Gaussian model from the results for a freely jointed chain. Because
Eq.~\eqref{eq:rg2decomposition} is a sum of positive contributions, the ratio
plotted in FIG.~\ref{fig:fig2} is normalized, and therefore the behavior seen in
FIG.~\ref{fig:fig2} expresses the swelling of these chains in water. Thus it is
clear that the quantitative discrepancies are significant, though these
characteristics offer minimal expression of molecular detail.

\begin{figure}[h]
\begin{center} 
\includegraphics[width=3.2in]{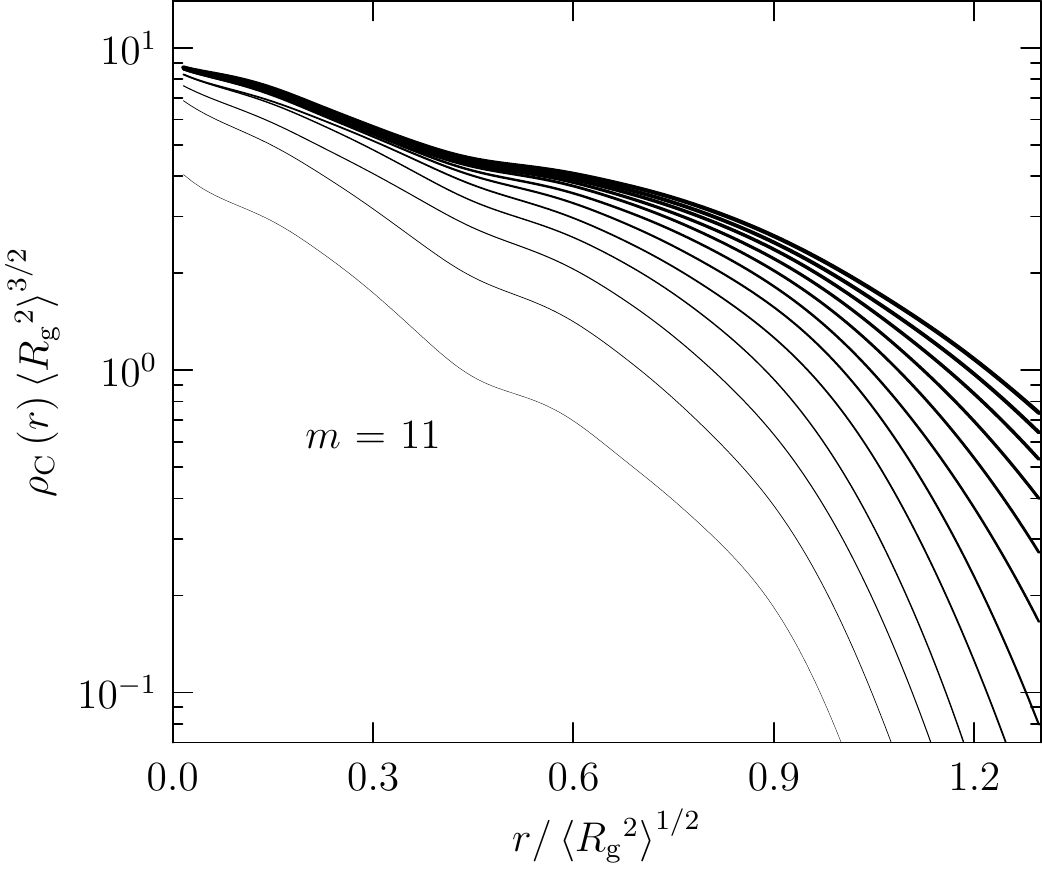}
\includegraphics[width=3.2in]{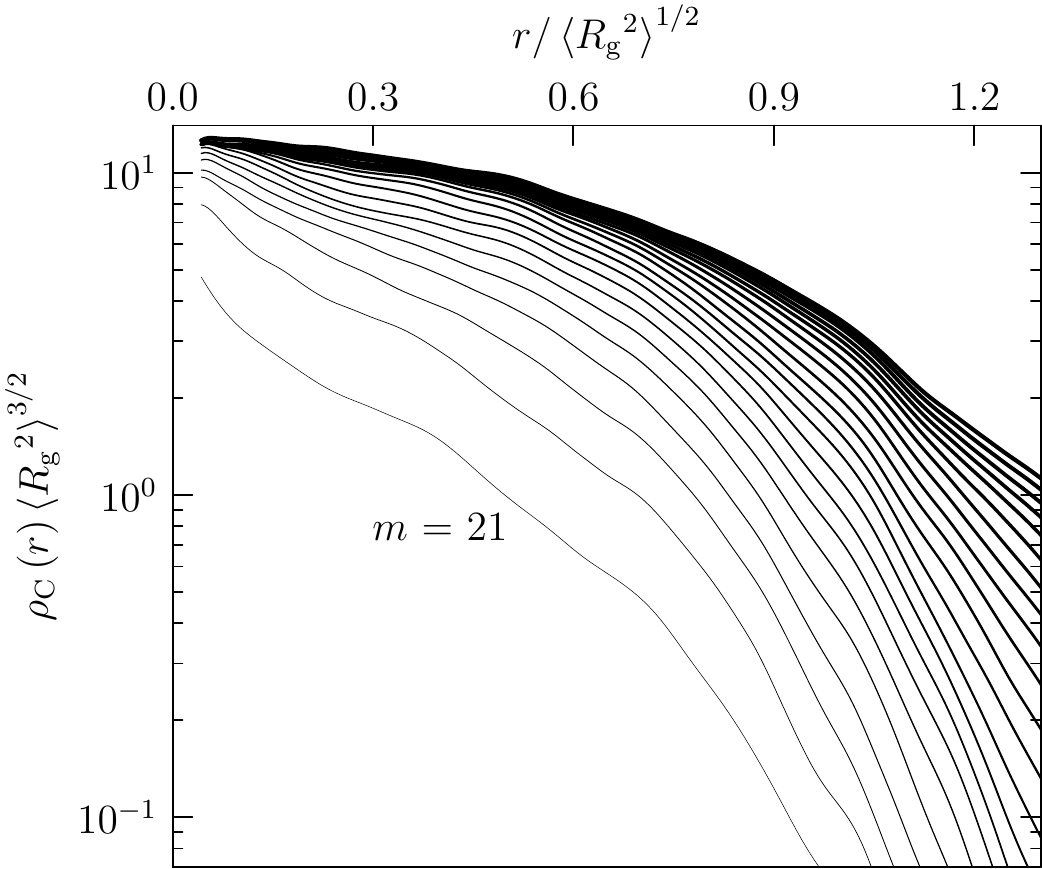}
\end{center} 
\caption{C atom contributions to the chain density profiles from their center of
mass, cumulatively indexed by the contour distance from the chain center. The
lowest curve corresponds to the mid-pair C atoms, the next higher curve includes
two additional C atoms, and the upper-most curve includes all C atoms.
\label{fig:fig3}} \end{figure}

Identifying chemically distinct C atoms permits a layered display of density
profiles (FIG.~\ref{fig:fig3}). C atoms near the center of the chain are likely
in the interior of the chain droplet, and the end-atoms are more likely on the
outside. For the $m$ = 11 chains some molecular-scale structure is evident, but
that is less prominent (FIG.~\ref{fig:fig3}) for $m$ = 21 chains.

The natural comparative model for the density profiles,\cite{YamakawaA}
\begin{eqnarray} \rho_{\mathrm{C}}\left(r\right) = \sum_{j = \mathrm{C~atoms}}
\frac{ \me^{- 3 r^2/ 2 \left\langle R_g{}^2(j)\right\rangle} }{ \sqrt[3]{2\pi
\left\langle R_g{}^2(j)\right\rangle/3}}~, 
\end{eqnarray} 
is obtained by
superposing of the Gaussian distributions associated with the observed
$\left\langle R_g{}^2(j)\right\rangle$. The density contributions of C atoms
interior the chains are more structured than the overall density profile.

\begin{figure}[h]
\begin{center} \includegraphics[width=3.2in]{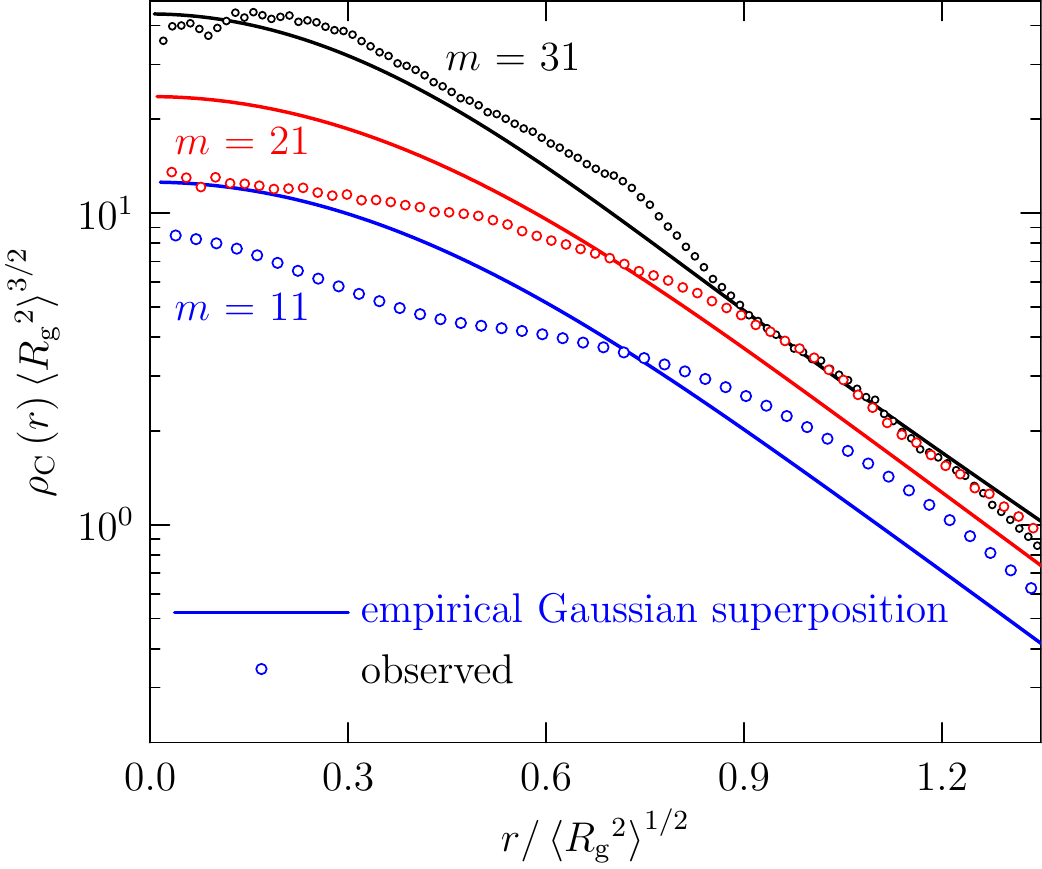}
\end{center} 
\caption{Density profiles for C atoms, relative to their center of mass. The
Gaussian superposition models (solid curves) follow from the discussion of
Yamakawa,\cite{YamakawaA} using the empirical $\left\langle
R_g{}^2(j)\right\rangle$ shown in FIG.~\ref{fig:fig1}. \label{fig:fig4}}
\end{figure}

\section{Conclusions} Chain structure factors, relevant to neutron scattering
from a chain in neutral water, are compared in detail to a traditional
continuum-Gaussian model result. The most serious limitation of the traditional
continuum-Gaussian structure factor is the failure to match the trivially known
$k \rightarrow \infty$ limiting value. A discrete-Gaussian model that is
consistent with the correct $k \rightarrow \infty$ value is considered.
Shifting-and-scaling the discrete-Gaussian model helps to identify the low-$k$
to high-$k$ transition near $k \approx 2\pi/0.6~\mathrm{nm}$ when an empirically
matched number of Gaussian links is about one-third of the total number of
effective-atom sites. The shifted-and-scaled discrete-Gaussian model better
identifies the transition from low-$k$ to high-$k$ behavior near $k \approx
2\pi/0.6\mathrm{nm}$, which thus provides a natural spatial size for the
coarsened monomers. This short distance-scale boundary of 0.6~nm is directly
verified with the $r$-space distributions.

Further testing of Gaussian chain models for these systems shows that
$\left\langle R_g{}^2(j)\right\rangle$, the contribution of the $j$-th chain
segment to $\left\langle R_g{}^2\right\rangle$, depends on contour index about
as expected for Gaussian chains despite quantitative discrepancies. The
quantitative comparison expresses the swelling of these chains in water.
Monomers central to the chain contour are usually central to the chain globule.
The density profiles of chain molecule segments relative to their center of mass
can show distinctive density structuring for smaller chains due close proximity
of central elements to the globule center. That density structuring washes-out
for longer chains, and due to the coarsened length-scale $\left\langle
R_\mathrm{g}{}^2\right\rangle^{1/2}$, many chain elements then contributing
additively to the density profiles. Gaussian chain models thus become more
satisfactory for the density profiles for longer chains. 

\section*{Acknowledgements}
The financial support of the Gulf of Mexico Research Initiative
(Consortium for Ocean Leadership Grant SA 12-05/GoMRI-002) is gratefully
acknowledged.

\bibliographystyle{achemso}

%\bibliography{ECapPairing,bib}

\begin{mcitethebibliography}{32}
\providecommand*\natexlab[1]{#1}
\providecommand*\mciteSetBstSublistMode[1]{}
\providecommand*\mciteSetBstMaxWidthForm[2]{}
\providecommand*\mciteBstWouldAddEndPuncttrue
  {\def\EndOfBibitem{\unskip.}}
\providecommand*\mciteBstWouldAddEndPunctfalse
  {\let\EndOfBibitem\relax}
\providecommand*\mciteSetBstMidEndSepPunct[3]{}
\providecommand*\mciteSetBstSublistLabelBeginEnd[3]{}
\providecommand*\EndOfBibitem{}
\mciteSetBstSublistMode{f}
\mciteSetBstMaxWidthForm{subitem}{(\alph{mcitesubitemcount})}
\mciteSetBstSublistLabelBeginEnd
  {\mcitemaxwidthsubitemform\space}
  {\relax}
  {\relax}

\bibitem[Alessi et~al.(2005)Alessi, Norman, Knowlton, Ho, and
  Greer]{Alessi:2005ix}
Alessi,~M.~L.; Norman,~A.~I.; Knowlton,~S.~E.; Ho,~D.~L.; Greer,~S.~C.
  \emph{Macromolecules} \textbf{2005}, \emph{38}, 9333--9340\relax
\mciteBstWouldAddEndPuncttrue
\mciteSetBstMidEndSepPunct{\mcitedefaultmidpunct}
{\mcitedefaultendpunct}{\mcitedefaultseppunct}\relax
\EndOfBibitem
\bibitem[Norman et~al.(2007)Norman, Fei, Ho, and Greer]{Norman:2007kq}
Norman,~A.~I.; Fei,~Y.; Ho,~D.~L.; Greer,~S.~C. \emph{Macromolecules}
  \textbf{2007}, \emph{40}, 2559--2567\relax
\mciteBstWouldAddEndPuncttrue
\mciteSetBstMidEndSepPunct{\mcitedefaultmidpunct}
{\mcitedefaultendpunct}{\mcitedefaultseppunct}\relax
\EndOfBibitem
\bibitem[dis(2005)]{dispersants}
\emph{Understanding Oil Spill Dispersants: Efficacy and Effects}; National
  Academies Press, Washington DC, 2005\relax
\mciteBstWouldAddEndPuncttrue
\mciteSetBstMidEndSepPunct{\mcitedefaultmidpunct}
{\mcitedefaultendpunct}{\mcitedefaultseppunct}\relax
\EndOfBibitem
\bibitem[Lin and Rubtsov(2012)Lin, and Rubtsov]{Lin:2012dz}
Lin,~Z.; Rubtsov,~I.~V. \emph{Proc. Nat. Acad. Sci. USA} \textbf{2012},
  \emph{109}, 1413--1418\relax
\mciteBstWouldAddEndPuncttrue
\mciteSetBstMidEndSepPunct{\mcitedefaultmidpunct}
{\mcitedefaultendpunct}{\mcitedefaultseppunct}\relax
\EndOfBibitem
\bibitem[Weikl(2008)]{Weikl:2008ii}
Weikl,~T.~R. \emph{Arch. Biochem. Biophys.} \textbf{2008}, \emph{469},
  67--75\relax
\mciteBstWouldAddEndPuncttrue
\mciteSetBstMidEndSepPunct{\mcitedefaultmidpunct}
{\mcitedefaultendpunct}{\mcitedefaultseppunct}\relax
\EndOfBibitem
\bibitem[Chaudhari et~al.(2010)Chaudhari, Pratt, and
  Paulaitis]{chaudhari_communication:_2010}
Chaudhari,~M.~I.; Pratt,~L.~R.; Paulaitis,~M.~E. \emph{J. Chem. Phys.}
  \textbf{2010}, \emph{133}, 231102\relax
\mciteBstWouldAddEndPuncttrue
\mciteSetBstMidEndSepPunct{\mcitedefaultmidpunct}
{\mcitedefaultendpunct}{\mcitedefaultseppunct}\relax
\EndOfBibitem
\bibitem[Chaudhari(2013)]{MICThesis}
Chaudhari,~M.~I. Molecular Simulations to Study Thermodynamics of Polyethylene
  Oxide Solutions. Ph.D.\ thesis, Department of Chemical {\&} Biomolecular
  Engineering, Tulane University, 2013\relax
\mciteBstWouldAddEndPuncttrue
\mciteSetBstMidEndSepPunct{\mcitedefaultmidpunct}
{\mcitedefaultendpunct}{\mcitedefaultseppunct}\relax
\EndOfBibitem
\bibitem[Dormidontova(2004)]{Dormidontova:2004uz}
Dormidontova,~E.~E. \emph{Macromolecules} \textbf{2004}, \emph{37},
  7747--7761\relax
\mciteBstWouldAddEndPuncttrue
\mciteSetBstMidEndSepPunct{\mcitedefaultmidpunct}
{\mcitedefaultendpunct}{\mcitedefaultseppunct}\relax
\EndOfBibitem
\bibitem[Nichols and Pratt(1982)Nichols, and Pratt]{nichols1982disentanglement}
Nichols,~A.~L.; Pratt,~L.~R. \emph{Faraday Symp. Chem. Soc.} \textbf{1982},
  \emph{17}, 129--140\relax
\mciteBstWouldAddEndPuncttrue
\mciteSetBstMidEndSepPunct{\mcitedefaultmidpunct}
{\mcitedefaultendpunct}{\mcitedefaultseppunct}\relax
\EndOfBibitem
\bibitem[Wilson et~al.(1984)Wilson, Nichols~III, and Pratt]{Wilson:1984cs}
Wilson,~M.~A.; Nichols~III,~A.~L.; Pratt,~L.~R. \emph{J. Chem. Phys.}
  \textbf{1984}, \emph{81}, 579--580\relax
\mciteBstWouldAddEndPuncttrue
\mciteSetBstMidEndSepPunct{\mcitedefaultmidpunct}
{\mcitedefaultendpunct}{\mcitedefaultseppunct}\relax
\EndOfBibitem
\bibitem[Nichols~III and Pratt(1984)Nichols~III, and Pratt]{nichols1984salt}
Nichols~III,~A.~L.; Pratt,~L.~R. \emph{J. Chem Phys.} \textbf{1984}, \emph{80},
  6225--6233\relax
\mciteBstWouldAddEndPuncttrue
\mciteSetBstMidEndSepPunct{\mcitedefaultmidpunct}
{\mcitedefaultendpunct}{\mcitedefaultseppunct}\relax
\EndOfBibitem
\bibitem[Pratt(1992)]{pratt1992contact}
Pratt,~L.~R. \emph{J. Phys. Chem.} \textbf{1992}, \emph{96}, 25--33\relax
\mciteBstWouldAddEndPuncttrue
\mciteSetBstMidEndSepPunct{\mcitedefaultmidpunct}
{\mcitedefaultendpunct}{\mcitedefaultseppunct}\relax
\EndOfBibitem
\bibitem[Borodin et~al.(2002)Borodin, Bedrov, and Smith]{Borodin:2002gx}
Borodin,~O.; Bedrov,~D.; Smith,~G.~D. \emph{Macromolecules} \textbf{2002},
  \emph{35}, 2410--2412\relax
\mciteBstWouldAddEndPuncttrue
\mciteSetBstMidEndSepPunct{\mcitedefaultmidpunct}
{\mcitedefaultendpunct}{\mcitedefaultseppunct}\relax
\EndOfBibitem
\bibitem[Israelachvili(1997)]{Israelachvili:1997tr}
Israelachvili,~J. \emph{Proc. Nat.l Acad. Sci. USA} \textbf{1997}, \emph{94},
  8378--8379\relax
\mciteBstWouldAddEndPuncttrue
\mciteSetBstMidEndSepPunct{\mcitedefaultmidpunct}
{\mcitedefaultendpunct}{\mcitedefaultseppunct}\relax
\EndOfBibitem
\bibitem[Cohen et~al.(2009)Cohen, Podgornik, Hansen, and
  Parsegian]{Cohen:2009ct}
Cohen,~J.~A.; Podgornik,~R.; Hansen,~P.~L.; Parsegian,~V.~A. \emph{J. Phys.
  Chem. B} \textbf{2009}, \emph{113}, 3709--3714\relax
\mciteBstWouldAddEndPuncttrue
\mciteSetBstMidEndSepPunct{\mcitedefaultmidpunct}
{\mcitedefaultendpunct}{\mcitedefaultseppunct}\relax
\EndOfBibitem
\bibitem[Cohen et~al.(2012)Cohen, Podgornik, and Parsegian]{Cohen:2012ul}
Cohen,~J.~A.; Podgornik,~R.; Parsegian,~V.~A. \emph{Biophys. J.} \textbf{2012},
  \emph{102}, 400A--400A\relax
\mciteBstWouldAddEndPuncttrue
\mciteSetBstMidEndSepPunct{\mcitedefaultmidpunct}
{\mcitedefaultendpunct}{\mcitedefaultseppunct}\relax
\EndOfBibitem
\bibitem[Bae et~al.(1993)Bae, Shim, and Prausnitz]{Bae:1993uj}
Bae,~Y.~C.; Shim,~D.~S.,~J. J .and~Soane; Prausnitz,~J.~M. \emph{J. Appl. Poly.
  Sci.} \textbf{1993}, \emph{47}, 1193--1206\relax
\mciteBstWouldAddEndPuncttrue
\mciteSetBstMidEndSepPunct{\mcitedefaultmidpunct}
{\mcitedefaultendpunct}{\mcitedefaultseppunct}\relax
\EndOfBibitem
\bibitem[Zafarani-Moattar and Tohidifar(2006)Zafarani-Moattar, and
  Tohidifar]{ZafaraniMoattar:2006kr}
Zafarani-Moattar,~M.~T.; Tohidifar,~N. \emph{J. Chem. Eng. Data} \textbf{2006},
  \emph{51}, 1769--1774\relax
\mciteBstWouldAddEndPuncttrue
\mciteSetBstMidEndSepPunct{\mcitedefaultmidpunct}
{\mcitedefaultendpunct}{\mcitedefaultseppunct}\relax
\EndOfBibitem
\bibitem[Zafarani-Moattar and Tohidifar(2008)Zafarani-Moattar, and
  Tohidifar]{ZafaraniMoattar:2008tb}
Zafarani-Moattar,~M.~T.; Tohidifar,~N. \emph{J. Chem. Eng. Data} \textbf{2008},
  \emph{53}, 785--793\relax
\mciteBstWouldAddEndPuncttrue
\mciteSetBstMidEndSepPunct{\mcitedefaultmidpunct}
{\mcitedefaultendpunct}{\mcitedefaultseppunct}\relax
\EndOfBibitem
\bibitem[Chaudhari and Pratt(2012)Chaudhari, and Pratt]{Chaudhari2014}
Chaudhari,~M.~I.; Pratt,~L.~R. In \emph{OIL SPILL REMEDIATION: COLLOID
  CHEMISTRY-BASED PRINCIPLES AND SOLUTIONS}; Somasundaran,~P., Farinato,~R.,
  Patra,~P., Papadopoulos,~K.~D., Eds.; John Wiley and Sons, Inc., 2012; See
  also: arXiv:1208.0349v2\relax
\mciteBstWouldAddEndPuncttrue
\mciteSetBstMidEndSepPunct{\mcitedefaultmidpunct}
{\mcitedefaultendpunct}{\mcitedefaultseppunct}\relax
\EndOfBibitem
\bibitem[Borodin et~al.(2001)Borodin, Bedrov, and Smith]{Borodin:2001cn}
Borodin,~O.; Bedrov,~D.; Smith,~G.~D. \emph{Macromolecules} \textbf{2001},
  \emph{34}, 5687--5693\relax
\mciteBstWouldAddEndPuncttrue
\mciteSetBstMidEndSepPunct{\mcitedefaultmidpunct}
{\mcitedefaultendpunct}{\mcitedefaultseppunct}\relax
\EndOfBibitem
\bibitem[Lee et~al.(2008)Lee, Venable, MacKerell~Jr, and Pastor]{Lee:2008hb}
Lee,~H.; Venable,~R.~M.; MacKerell~Jr,~A.~D.; Pastor,~R.~W. \emph{Biophys. J.}
  \textbf{2008}, \emph{95}, 1590--1599\relax
\mciteBstWouldAddEndPuncttrue
\mciteSetBstMidEndSepPunct{\mcitedefaultmidpunct}
{\mcitedefaultendpunct}{\mcitedefaultseppunct}\relax
\EndOfBibitem
\bibitem[Choi et~al.(2013)Choi, Mondal, and Yethiraj]{Choi:2013dg}
Choi,~E.; Mondal,~J.; Yethiraj,~A. \emph{J. Phys. Chem. B} \textbf{2013},
  131218200604003\relax
\mciteBstWouldAddEndPuncttrue
\mciteSetBstMidEndSepPunct{\mcitedefaultmidpunct}
{\mcitedefaultendpunct}{\mcitedefaultseppunct}\relax
\EndOfBibitem
\bibitem[Mondal et~al.(2014)Mondal, Choi, and Yethiraj]{Mondal:2014gp}
Mondal,~J.; Choi,~E.; Yethiraj,~A. \emph{Macromolecules} \textbf{2014},
  \emph{47}, 438--446\relax
\mciteBstWouldAddEndPuncttrue
\mciteSetBstMidEndSepPunct{\mcitedefaultmidpunct}
{\mcitedefaultendpunct}{\mcitedefaultseppunct}\relax
\EndOfBibitem
\bibitem[Starovoytov et~al.(2011)Starovoytov, Borodin, Bedrov, and
  Smith]{Starovoytov:2011km}
Starovoytov,~O.~N.; Borodin,~O.; Bedrov,~D.; Smith,~G.~D. \emph{J. Chem. Theory
  Comput.} \textbf{2011}, \emph{7}, 1902--1915\relax
\mciteBstWouldAddEndPuncttrue
\mciteSetBstMidEndSepPunct{\mcitedefaultmidpunct}
{\mcitedefaultendpunct}{\mcitedefaultseppunct}\relax
\EndOfBibitem
\bibitem[Earl and Deem(2005)Earl, and Deem]{Earl:2005fv}
Earl,~D.~J.; Deem,~M.~W. \emph{Phys. Chem. Chem. Phys.} \textbf{2005},
  \emph{7}, 3910--3916\relax
\mciteBstWouldAddEndPuncttrue
\mciteSetBstMidEndSepPunct{\mcitedefaultmidpunct}
{\mcitedefaultendpunct}{\mcitedefaultseppunct}\relax
\EndOfBibitem
\bibitem[Jorgensen et~al.(1996)Jorgensen, Maxwell, and
  Tirado-Rives]{Jorgensen:1996vx}
Jorgensen,~W.~L.; Maxwell,~D.~S.; Tirado-Rives,~J. \emph{J. Am. Chem. Soc.}
  \textbf{1996}, \emph{118}, 11225--11236\relax
\mciteBstWouldAddEndPuncttrue
\mciteSetBstMidEndSepPunct{\mcitedefaultmidpunct}
{\mcitedefaultendpunct}{\mcitedefaultseppunct}\relax
\EndOfBibitem
\bibitem[Berendsen et~al.(1987)Berendsen, Grigera, and
  Straatsma]{Berendsen:1987uu}
Berendsen,~H. J.~C.; Grigera,~J.~R.; Straatsma,~T.~P. \emph{J. Phys. Chem.}
  \textbf{1987}, \emph{91}, 6269--6271\relax
\mciteBstWouldAddEndPuncttrue
\mciteSetBstMidEndSepPunct{\mcitedefaultmidpunct}
{\mcitedefaultendpunct}{\mcitedefaultseppunct}\relax
\EndOfBibitem
\bibitem[van~der Spoel et~al.(2005)van~der Spoel, Lindahl, Hess, Groenhof,
  Mark, and Berendsen]{vanderSpoel:2005hz}
van~der Spoel,~D.; Lindahl,~E.; Hess,~B.; Groenhof,~G.; Mark,~A.~E.;
  Berendsen,~H. J.~C. \emph{J. Comp. Chem.} \textbf{2005}, \emph{26},
  1701--1718\relax
\mciteBstWouldAddEndPuncttrue
\mciteSetBstMidEndSepPunct{\mcitedefaultmidpunct}
{\mcitedefaultendpunct}{\mcitedefaultseppunct}\relax
\EndOfBibitem
\bibitem[Berne and Pecora(1976)Berne, and Pecora]{BernePecora}
Berne,~B.~J.; Pecora,~R. \emph{Dynamic Light Scattering}; John Wiley \& Sons:
  New York, 1976\relax
\mciteBstWouldAddEndPuncttrue
\mciteSetBstMidEndSepPunct{\mcitedefaultmidpunct}
{\mcitedefaultendpunct}{\mcitedefaultseppunct}\relax
\EndOfBibitem
\bibitem[Yamakawa(1971)]{YamakawaB}
Yamakawa,~H. \emph{Modern Theory of Polymer Solutions}; Harper \& Row: New
  York, 1971; Eq. (8.50)\relax
\mciteBstWouldAddEndPuncttrue
\mciteSetBstMidEndSepPunct{\mcitedefaultmidpunct}
{\mcitedefaultendpunct}{\mcitedefaultseppunct}\relax
\EndOfBibitem
\bibitem[Yamakawa(1971)]{YamakawaA}
Yamakawa,~H. \emph{Modern Theory of Polymer Solutions}; Harper \& Row: New
  York, 1971; Sec. 7a\relax
\mciteBstWouldAddEndPuncttrue
\mciteSetBstMidEndSepPunct{\mcitedefaultmidpunct}
{\mcitedefaultendpunct}{\mcitedefaultseppunct}\relax
\EndOfBibitem
\end{mcitethebibliography}

\providecommand*\mcitethebibliography{\thebibliography}
\csname @ifundefined\endcsname{endmcitethebibliography}
  {\let\endmcitethebibliography\endthebibliography}{}

\end{document}